\documentclass[twocolumn,showpacs,preprintnumbers,draftclsnofoot]{revtex4}
\usepackage{amsmath}
\usepackage{amssymb}
\usepackage{graphicx}
\usepackage{dcolumn}
\usepackage{bm}
\usepackage{amssymb}
\usepackage{graphicx}
\usepackage{dcolumn}
\usepackage{bm}

\begin{document}

\title{ Controlling the Interferometers of Zero-Line Modes in Graphene by Pseudomagnetic field   }
\author{Ma Luo\footnote{Corresponding author:luom28@mail.sysu.edu.cn} }
\affiliation{The State Key Laboratory of Optoelectronic Materials and Technologies \\
School of Physics\\
Sun Yat-Sen University, Guangzhou, 510275, P.R. China}

\begin{abstract}

Networks of graphene-based topological domain walls function as nano-scale interferometers of zero-line modes, with magnetic field and(or) scalar potential as the controlling parameters. In the absence of externally applied magnetic or electrical field, strain induces pseudomagnetic field and scalar potential in graphene, which could control the interferometers more efficiently. Two types of strains are considered: (i) Horizontally bending the graphene nanoribbon into circular arc induces nearly uniform pseudomagnetic field; (ii) Helicoidal graphene nanoribbon exhibit nonuniform pseudomagnetic field. Both types of strain induce small scalar potential due to dilatation. The interferometers are studied by transport calculation of the tight binding model. The transmission rates through the interferometer depend on the strain parameters. An interferometer with three loops is designed, which could completely switch the transmitting current from one export to the other.

\end{abstract}

 \maketitle

\section{Introduction}

\begin{figure}[tbp]
\scalebox{0.56}{\includegraphics{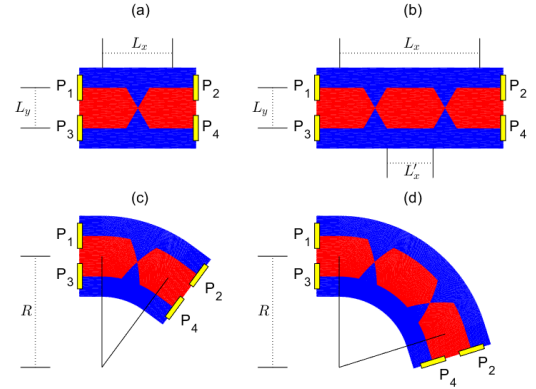}}
\caption{ The spatial structure of the ZLM splitter and AB interferometer base on network of domain walls in graphene nanoribbon. The free standing heterostructure is plotted at the top, and the horizontally bent heterostructure is plotted at the bottom. The regions with $\eta_{i}=+1(-1)$ are filled with red(blue) color. The four ports to the four corresponding domain walls are marked by yellow pads. The geometry parameters of the interferometer are marked by $L_{x}$, $L_{x}^{\prime}$ and $L_{y}$; the bending radius is $R$. \label{fig1}}
\end{figure}

Graphene based nano-structures with localized conductive quantum states make the integrated electronic and spintronic feasible. Topological zero-line modes (ZLMs) are robust localized conductive states that appear at the topological domain walls \cite{Semenoff08,Martin08,Zarenia11,Klinovaja12,Zarenia12,Abolhassan13,FanZhang13,Xintao15,Changhee16,YafeiRen17,TaoHou18}. One type of ZLMs are hosted at the domain wall of monolayer graphene between two regions with opposite staggered sublattice potential \cite{Semenoff08,Zarenia12}. The staggered sublattice potential is induced by h-BN or SiC substrate \cite{Giovannetti07,CRDean10,SYZhou07}. At the intersection between four domain walls [ZLMs splitter in Fig. \ref{fig1}(a)], the imported current from one of the domain wall is partitioned and exported into the two adjacent domain walls \cite{ZhenhuaQiao11,ZhenhuaQiao14,Anglin17,Wang17,YafeiRen17}. Because the inter-valley scattering is absent, all current at import and exports are carried by the ZLMs in the same band valley. For the third exporting domain walls, the ZLMs in the same band valley travels along inward direction, so that no currents is guided to this export. A network of ZLMs with one closed loop and two intersections along the loop form an interferometer, as shown in Fig. \ref{fig1}(b). In the presence of magnetic flux through the closed loop, the Aharanov-Bohm (AB) effect modifies the transmission rate between two external ports \cite{ShuguangCheng18}. For the closed loop with size of nanometer scale, a magnetic field around one Tesla is required to induce more than one quantized magnetic flux.

\begin{figure}[tbp]
\scalebox{0.58}{\includegraphics{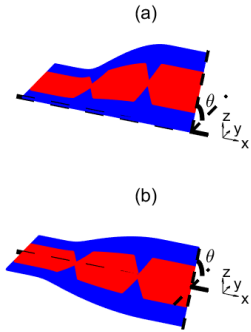}}
\caption{ The spatial structure of the AB interferometer in helicoidal graphene nanoribbon. The nanoribbon is twisted around the open boundary in (a), and the axis in (b), by the angle of $\theta$. \label{fig_HeliConfig}}
\end{figure}

In order to make the interferometer more feasible in integrated devices with smaller size, we proposed to replace the magnetic field by pseudomagnetic field that is induced by strain \cite{Guinea10,ZhigangWang13,ZenanQi14,ShuzeZhu15,Settnes16}. The presence of strain in graphene change the bond length, which is equivalent to a pseudomagnetic field. The pseudomagnetic field change the geometry phases of the ZLMs in the two arms of the interferometer. In addition, scalar potential is induced due to dilatation. Because the graphene is gapped by the staggered sublattice potential, the scalar potential is not screened. The scalar potential changes the dynamical phases of the ZLMs in the two arms of the interferometer. Combination of the phase changes exhibit interference pattern of the transmission rate. Two types of strain are considered: (i) Bending a graphene nanoribbon horizontally into circle [as shown in Fig. \ref{fig1}(c) and (d)] induces pseudomagnetic field that is nearly uniform\cite{Guinea10}. (ii) Twisting the graphene around the open boundary (or the axis) of the nanoribbon into helicoid  [as shown in Fig. \ref{fig_HeliConfig}(a) and (b)] induces nonuniform pseudomagnetic field and scalar potential. The transmission rates through the interferometer were numerically calculated and discussed.

The design of triple interferometer that consists of two interferometers in parallel enable the complete switching of the transmitting current from one export to the other. Because the strain as small as 1\% could generate pseudomagnetic field as large as 10 T, the complete switching of incident current requires small strain.

The article is organized as following: Section II described the system and theoretical model. Section III presents and discusses the numerical results of strain dependent transmission rate. Section IV describe the design of the triple interferometer. Section V is the conclusion.

\section{System and Model}

Zigzag nanoribbons of monolayer graphene with staggered sublattice potential are studied. The Hamiltonian of the tight binding model is given as
\begin{equation}
H=\sum_{\langle i,j\rangle}t_{ij}c_{i}^{\dag}c_{j}+\sum_{i}[\eta_{i}\Delta_{i}+V(\mathbf{r}_{i})]c_{i}^{\dag}c_{i}
\end{equation}
where $t_{ij}$ is the hopping parameter between the nearest neighbor sites $\langle i,j\rangle$, $\Delta_{i}=\pm\Delta_{0}$ if the $i$-th site belongs to A(B) sublattice, $\eta_{i}=\pm1$ if the $i$-th site belongs to the region with positive(negative) staggered sublattice potential, $V(\mathbf{r}_{i})$ is the scalar potential due to dilatation and external gated pads. For graphene without strain, the hopping parameter is $t_{0}=-2.7$ eV. We considered the systems with $\Delta_{0}=0.1t_{0}$ as example. The spatial structures of the ZLM splitter and AB interferometer in the nanoribbons are shown in Fig. \ref{fig1}. At the intersections of the splitters, the crossing angle between two importing domain walls is $2\pi/3$. Connecting two ZLM splitters in series form an AB interferometer. In our numerical calculation, we used the parameters that $L_{x}=23.61$ nm in (a), and $L_{x}=47.22$ nm, $L_{x}^{\prime}=15.74$ nm in (b); $L_{y}=13.63$ nm for both of the ZLM splitter and AB interferometer. The four leads that connect to the four external domain walls are plotted as yellow pads. The transmission rate $T_{i,j}$ with import from the $i$-th lead and export to the $j$-th lead is numerically calculated by non-equilibrium Green's function method \cite{LopezSancho84,Nardelli98,FufangXu07,Diniz12,Lewenkopf13}.

In the presence of strain, the coordinate of each lattice site $\mathbf{r}_{i}$ is changed to $\mathbf{r}_{i}+\mathbf{u}_{i}$, so that the bond length between nearest neighbor sites is changed. The hopping parameter is modified as \cite{ZenanQi14,Settnes16}
\begin{equation}
t_{ij}=t_{0}e^{-\beta(\frac{d_{ij}}{a_{c}}-1)}
\end{equation}
where $\beta=3.37$, $a_{c}=0.142$ nm is the bond length of unstrained graphene, $d_{ij}$ is the distance between the nearest neighbor sites $\langle i,j\rangle$ of the graphene with strain. For the low energy excitation near to the $K$ point of the Brillouin zone, the effective vector potential of the pseudomagnetic field at the i-th lattice site could be estimated as
\begin{equation}
A_{x}(\mathbf{r}_{i})-iA_{y}(\mathbf{r}_{i})=\frac{2\hbar}{3t_{0}a_{c}e}\sum_{j\in\langle i,j\rangle}(t_{ij}-t_{0})e^{i\mathbf{K}\cdot(\mathbf{r}_{j}-\mathbf{r}_{i})} \label{pseudoAvector}
\end{equation}
The dilatation of the graphene is given as $\nabla\cdot\mathbf{u}$, and the scalar potential is given as $V_{0}\nabla\cdot\mathbf{u}$ with $V_{0}\approx3$ eV \cite{Guinea10}.

For the systems with the first type of strain, the spatial structures are shown in Fig. \ref{fig1}(c) and (d). The scattering regions that contain the ZLM splitter or the AB interferometer [the regions within the range of $L_{x}$ in Fig. \ref{fig1}(a) and (b)] are bent into circle; the domain walls that connect to the leads remain straight. Assuming that the axial length of the nanoribbon is not changed by the bending, the bending angle is $\theta=L_{x}/R$, with $R$ being the bending radius. With small bending angle, the pseudomagnetic field is nearly uniform with strength $B_{s}\approx\frac{\Phi_{0}}{a_{c}R}$. The scalar potential is given as $V(\mathbf{r})\approx V_{0}y/R$, with $y$ being the coordinate of the lattice site in the corresponding unstrained graphene. If the bending angle is large, the pseudomagnetic field become nonuniform, and the scalar potential has more complicated form. The exact formula of the pseudomagnetic field and scalar potential can be found in Ref \cite{Guinea10}.


For the systems with the second type of strain, the spatial structures are shown in Fig. \ref{fig_HeliConfig}(a) and (b). The helicoidal graphene nanoribbon is characterized by the twisting angle $\theta$. Only the scattering regions [the regions within the range of $L_{x}$ in Fig. \ref{fig1}(a) and (b)] are twisted, and the remaining regions are unstrained. The pseudomagnetic field could be numerically calculated by Eq. (\ref{pseudoAvector}). The dilatation is calculated by measuring the expansion of the surface area of the helicoidal surface, which is given as
\begin{equation}
V(\mathbf{r})=V_{0}(\sqrt{1+\frac{y^{2}\theta^{2}}{4\pi^{2}L_{x}^{2}}}-1)
\end{equation}

\section{Numerical results}

The interfere pattern of the AB interferometer with pseudomagnetic field is different from that with real magnetic field. For the AB interferometer with real magnetic field, the partition ratio of each ZLM splitter, $\alpha$, is independent of the magnetic field; the magnetic field is assumed to be uniform, so the AB phases of the two arms, $\phi$, are the same; the scalar potential is absent, so the dynamical phases of the two arms, $k_{\parallel}l_{0}$, are the same. Adopting the transfer matrix method, the transmission rates are given as \cite{ShuguangCheng18}
\begin{equation}
T_{1,2}=\frac{(1-\alpha)^2}{(1-\alpha)^2+4\alpha\cos^{2}\frac{\phi+k_{\parallel}l_{0}}{2}} \label{T12B}
\end{equation}
, $T_{1,3}=1-T_{1,2}$ and $T_{1,4}=0$, where $k_{\parallel}$ is the wave number of the ZLM along the domain wall and $l_{0}$ is the length of each arm. By contrast, for the AB interferometer with pseudomagnetic field, $\alpha$ are dependent on the bending angle; $\phi$ of the two arms are different because the pseudomagnetic field is in general nonuniform; $k_{\parallel}l_{0}$ of the two arms are different due to the scalar potential. In subsection (A), the AB interferometers consisted of horizontally bent nanoribbons are studied in detail to compare with the AB interferometers with real magnetid field. We firstly studied the dependence of the transmission rate through the ZLM splitter on the pseudomagnetic field, and secondly studied the transmission rate through the AB interferometer. In subsection (B), the numerical result of the AB interferometers consisted of helicoidal nanoribbon is presented.

\subsection{Horizontally Bent Interferometer}

\begin{figure}[tbp]
\scalebox{0.53}{\includegraphics{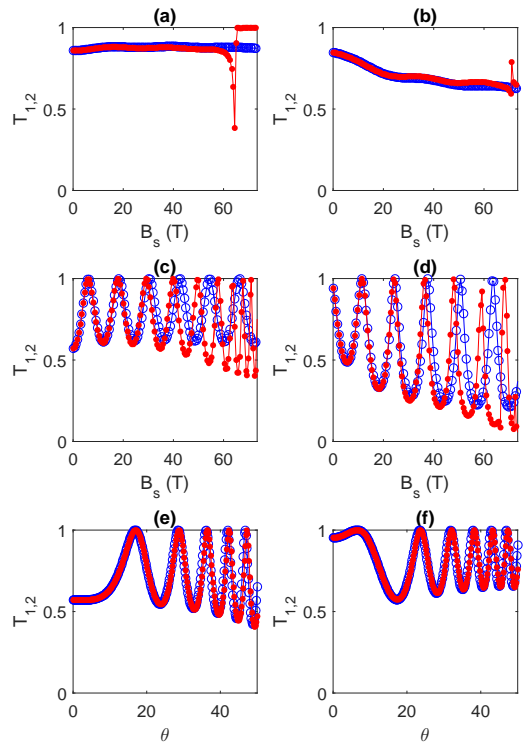}}
\caption{ The transmission rate between the $P_{1}$ and $P_{2}$ exports of the ZLM splitter and the AB interferometer versus the pseudomagnetic field $B_{s}$ in (a-b) and (c-d), respectively, for the nanoribbons that are horizontally bent. The transmission rate between the $P_{1}$ and $P_{2}$ exports of the interferometer versus the twisting angle $\theta$ in (e-f), for the nanoribbons that are twisted into helicoid. The Fermi energy level in (a,c,e) is $0$ eV, and that in (b,d,f) is $0.1$ eV. The numerical results that neglect or consider the scalar potential are plotted as blue(empty) dots or red(solid) dots, respectively. The solid lines are guide for eye.   \label{fig2}}
\end{figure}

The transmission rates through a single ZLM splitter versus the pseudomagnetic field are plotted in Fig. \ref{fig2}(a) and (b), where the Fermi energy equate to zero and $0.1$ eV, respectively. Because of the conservation of the valley index, the transmission rates from the $P_{1}$ port to the other three ports satisfy the conditions $T_{1,4}=0$ and $T_{1,3}=1-T_{1,2}$. Thus, only $T_{1,2}$ are plotted. The calculation results that neglect and consider the scalar potential are plotted together for comparison. The scalar potential is proportional to the bending angle, which in turn is proportional to the pseudomagnetic field. If the scalar potential is smaller than the energy difference between the Fermi energy and the conduction(valence) bulk band edge, only the dynamical phases of the ZLMs are changed. In contrast, if the scalar potential is large enough, the strong coupling between the ZLMs and the bulk states induces reflection, which interferes with the ZLM splitter and changes the transmission rate. As a result, if the pseudomagnetic field is larger than 60 T, the transmission rates are significantly impacted by the scalar potential. As the pseudomagnetic field varies within the range of $0\sim60$ T, the transmission rates smoothly change. The change is more dramatic for the systems with larger Fermi energy. These phenomenons can be explained by inspecting the mechanism of the current partition at the intersection \cite{ZhenhuaQiao14}. The current partition originates from the coupling of the incoming ZLM to the two adjacent outgoing ZLMs. With a given distance from the intersection, the coupling strength is determined by the lateral separations between the adjacent ZLMs. The two outgoing ZLMs have different angle from the incoming ZLM, thus have different lateral separations, so the transmission rates to the two outgoing ZLMs are different. In the presence of strain, the lateral separations are changed, so the coupling strength is modified. Thus, the transmission rates to the two outgoing ZLM are changed. For the ZLM with larger Fermi energy, the wave number along the ZLM, $k_{\parallel}$, is larger. Thus, the coupling between adjacent ZLMs occur in a shorter distance, $l_{c}\approx1/k_{\parallel}$. As a result, the transmission rate is more sensitive to the change of lateral separations as the Fermi energy increases.

The transmission rates through the AB interferometer are plotted in Fig. \ref{fig2}(c) and (d), where the Fermi energy equate to zero and $0.1$ eV, respectively. If the scalar potential is neglected, the interfere patterns are trigonometric functions with varying magnitude. As comparison, the interfere patterns given by Eq. (\ref{T12B}) are trigonometric functions with fixed magnitude. This difference is due to the dependence of $\alpha$ on the pseudomagnetic field. If the scalar potential is considered, the interfere patterns of the transmission rates oscillate with higher frequency as the pseudomagnetic field increases, because the relative difference between the dynamical phases of the ZLMs along the two arms is enlarged by the scalar potential. This effect is negligible as the pseudomagnetic field remains small. The transmission rate could be controlled by external scalar potential barrier along one of the two arms of the interferometer. If ferromagnetic insulating material is deposited on top of the domain wall, spin dependent potential barrier is induced \cite{Yokoyama08,Haugen08}. Thus, the systems could exhibit spin pumping effect \cite{Inoue16}.

\subsection{Helicoidal Interferometer}

For the AB interferometers consisted of helicoidal nanoribbon in Fig. \ref{fig_HeliConfig}(a), the numerical results of the transmission rate are plotted in Fig. \ref{fig2}(e,f). If the scalar potential is neglected, the interference pattern have negligible change. The oscillation frequency become larger as $\theta$ increases. For the system in Fig. \ref{fig_HeliConfig}(b), the pseudomagnetic field is odd function of $y$, so that the magnetic flux through the loop is zero; the scalar potential at the two arms are the same. Thus, no interference pattern would be exhibited.

\section{Triple Interferometer as current switch}

\begin{figure}[tbp]
\scalebox{0.33}{\includegraphics{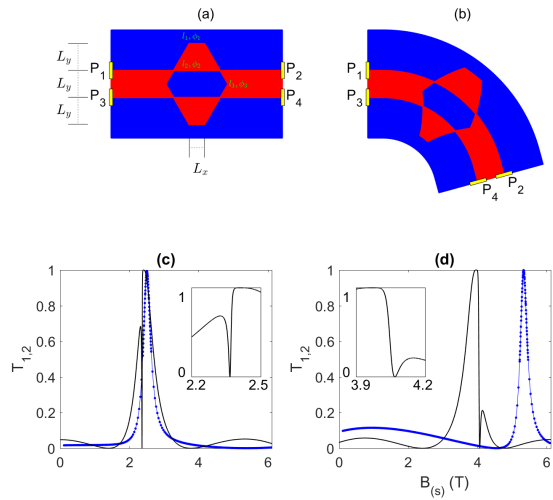}}
\caption{ Configuration of the triple interferometer in (a), and the bending of the nanostructure in (b). Transmission rate between the $P_{1}$ and $P_{2}$ exports versus pseudomagnetic field $B_{s}$(blue dots) or real magnetic field $B$(black solid lines). The Fermi energy level in (c) is $0$ eV, and that in (d) is $0.1$ eV.  \label{fig3}}
\end{figure}

Although the AB interferometer induces interfere pattern of the transmission rates, the nano-structure does not completely switch the imported current from one export to the other. More precisely, $T_{1,3}$ could equate to zero as the dynamical phase satisfy $(\phi+k_{\parallel}l_{0})/2=\pi(1/2+N)$ with $N$ being integer, but $T_{1,2}$ could never equate to zero as shown in Eq. (\ref{T12B}). We designed the triple interferometer in Fig. \ref{fig3}(a), which consists of two AB interferometers in parallel. The structure parameters are shown in Fig. \ref{fig3}(a), with $L_{x}=7.87$ nm and $L_{y}=13.63$ nm as example.


In the presence of uniform real magnetic field, the transmission rates to each export can be deduced by applying the scattering matrix method \cite{ShuguangCheng18}. In the network of ZLMs in Fig. \ref{fig3}(a), the length of each domain wall are marked as $l_{i}$. Path integral of the vector potential along each domain wall is marked as $\phi_{i}$. It is convenient to denote the round-trip phase of the upper and lower loops as $\Phi_{1}=k_{\parallel}l_{1}+k_{\parallel}l_{2}+\phi_{1}-\phi_{2}$, and the round-trip phase of the middle loop as $\Phi_{2}=2k_{\parallel}l_{2}+2k_{\parallel}l_{3}-2\phi_{2}-2\phi_{3}$. The transmission amplitudes to the exports are given as
\begin{widetext}
\begin{equation}
t_{1,2}=
\frac{-2(\alpha-1)[\cos(\Phi_{2})+\alpha \cos(\Phi_{1})]\cos(\frac{\Phi_{2}}{2})e^{i(k_{\parallel}l_{1}+k_{\parallel}l_{2}+k_{\parallel}l_{3}+\phi_{1}+\phi_{2}+\phi_{3})}}{(\alpha-1)^{2}e^{2i(k_{\parallel}l_{2}+k_{\parallel}l_{3})}+e^{2i\phi_{3}}(\alpha e^{i(k_{\parallel}l_{1}+k_{\parallel}l_{2}+\phi_{1})}+e^{i\phi_{2}})^{2}} \label{T12switch}
\end{equation}
\begin{equation}
t_{1,3}=\frac{i\alpha e^{i(k_{\parallel}l_{3}+\phi_{3}+2\phi_{3})}(e^{i\Phi_{1}}+1)^{2}}{(\alpha-1)^{2}e^{2i(k_{\parallel}l_{2}+k_{\parallel}l_{3})}+e^{2i\phi_{3}}(\alpha e^{i(k_{\parallel}l_{1}+k_{\parallel}l_{2}+\phi_{1})}+e^{i\phi_{2}})^{2}} \label{T13switch}
\end{equation}
\end{widetext}
The transmission rates are $T_{1,2}=|t_{1,2}|^{2}$ and $T_{1,3}=|t_{1,3}|^{2}$. The transmission rates are plotted as black(solid) lines in Fig. \ref{fig3}(c-d). The resonant peak of $T_{1,2}$ consists of a wide peak and a sharp dip, as shown in Fig. \ref{fig3}(c-d). The inserts in Fig \ref{fig3}(c-d) show that the sharp dip has asymmetric line shape, which implies a Fano type of resonance. The peak with $T_{1,2}=1$ is determined by the condition $\Phi_{1}=\pi(1+2N)$ with $N$ being integer. The sharp dip with $T_{1,2}=0$ is determined by the condition $\cos(\Phi_{2})+\alpha \cos(\Phi_{1})=0$. The sharp dip is due to the interfere between the two tunneling processes, i.e. the near-resonant tunneling through the upper and lower loops with round-trip phases being $\Phi_{1}$ and the off-resonant tunneling through the middle loop with round-trip phase being $\Phi_{2}$. The transmission valley in Fig \ref{fig3}(c-d) with $T_{1,2}=0$ is determined by the condition $\Phi_{2}=\pi(1+2N)$. In summary, the triple interferometer can completely switch the current from one export to the other by changing the real magnetic field.

\begin{figure}[tbp]
\scalebox{0.58}{\includegraphics{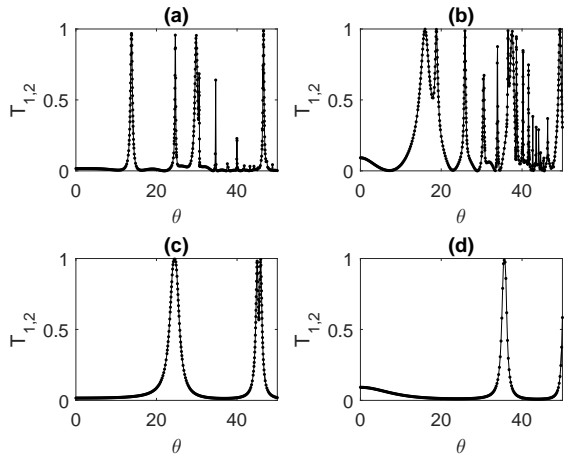}}
\caption{ Configuration of the triple interferometer in (a), and the bending of the nanostructure in (b). Transmission rate between the $P_{1}$ and $P_{2}$ exports versus pseudomagnetic field $B_{s}$(blue dots) or real magnetic field $B$(black solid lines). The Fermi energy level in (c) is $0$ eV, and that in (d) is $0.1$ eV.  \label{fig4}}
\end{figure}

If the nanoribbon is horizontally bent, as shown in Fig. \ref{fig3}(b), the pseudomagnetic field control the transmission rates, as shown by the numerical result in Fig. \ref{fig3}(c-d). The transmission rates are different from Eq. (\ref{T12switch}) and (\ref{T13switch}) due to the same reasons as those for the AB interferometers. With energy $\varepsilon=0$, the resonant peak with $T_{1,2}=1$ is slightly different from the peak given by Eq. (\ref{T12switch}), as shown in Fig. \ref{fig3}(c). The sharp dip due to Fano resonance vanishes. With energy $\varepsilon\neq0$, the resonant peak with $T_{1,2}=1$ is significantly different from the peak given by Eq. (\ref{T12switch}), as shown in Fig. \ref{fig3}(d), because the dynamical phase is more relevant. The transmission valleys with $T_{1,2}=0$ also appear within the range of the pseudomagnetic field in Fig. \ref{fig3}(c-d). Thus, the triple interferometer can completely switch the transmitting current from one export to the other. The largest pseudomagnetic field being considered in Fig. \ref{fig3} is $6$ T, which corresponds to bending angle as small as $4.80^{\circ}$ and maximum stretch of bond length as small as $2\%$. As a result, the bending of the nanoribbon is not as exaggerated as being shown in Fig. \ref{fig3}(b).

If the nanoribbon is twisted into helicoid, the transmission rate has interference patterns that are significantly different from Eq. (\ref{T12switch}) and (\ref{T13switch}), because the pseudomagnetic field is nonuniform. If the twisting is around the open boundary of the nanoribbon [similar to that in Fig. \ref{fig_HeliConfig}(a)], the transmission rate is plotted in Fig. \ref{fig4}(a) and (b). Because the pseudomagnetic field along all of the four arms are different, the interference patterns include many resonant peaks. As energy increase, more resonant peaks appear. If the twisting is around the axis of the nanoribbon [similar to that in Fig. \ref{fig_HeliConfig}(b)], the transmission rates are shown in Fig. \ref{fig4}(c) and (d). Because the pseudomagnetic field at the upper two arms has the same magnitude as that at the lower two arms, the interfere patterns have much fewer resonant peaks.

\section{conclusion}

In conclusion, quantum interferometers based on network of domain walls and the presence of pseudomagnetic field are theoretically studied. The strain in the horizontally bent graphene nanoribbon or helicoidal graphene nanoribbon induces pseudomagnetic field and scalar potential. In addition to inducing the AB phases, the strain changes the dynamical phases of the ZLMs along the domain walls and the partition ratio of each interaction. As a result, comparing to the interferometer with real uniform magnetic field, the interferometer with pseudomagnetic field has more complicated interfere pattern. The benefit of designing the interferometer with pseudomagnetic field is that large pseudomagnetic field can be obtained by small strain. Triple interferometers are designed to completely switch the transmission of current between two exports. These designs could be implemented experimentally and be developed into practical highly integrated nano-electronic devices.

\begin{acknowledgments}
The project is supported by the National Natural Science Foundation of China (Grant:
11704419), the National Basic Research
Program of China (Grant: 2013CB933601), and the National Key Research and Development Project of China
(Grant: 2016YFA0202001).
\end{acknowledgments}

\clearpage


\section*{References}

\begin{thebibliography}{99}


\bibitem{Semenoff08} G. W. Semenoff, V. Semenoff, and Fei Zhou, Phys. Rev. Lett. 101, 087204(2008).

\bibitem{Martin08} Ivar Martin, Ya. M. Blanter, and A. F. Morpurgo, Phys. Rev. Lett. 100, 036804(2008).

\bibitem{Zarenia11} M. Zarenia, J. M. Pereira, Jr., G. A. Farias, and F. M. Peeters, Phys. Rev. B 84, 125451(2011).

\bibitem{Klinovaja12} Jelena Klinovaja, Gerson J. Ferreira, and Daniel Loss, Phys. Rev. B 86, 235416(2012).

\bibitem{Zarenia12} M. Zarenia, O. Leenaerts, B. Partoens, and F. M. Peeters, Phys. Rev. B 86, 085451(2012)

\bibitem{Abolhassan13} Abolhassan Vaezi, Yufeng Liang, Darryl H. Ngai, Li Yang, and Eun-Ah Kim, Phys. Rev. X 3, 021018(2013).

\bibitem{FanZhang13} Fan Zhang, Allan H. MacDonald, and Eugene J. Mele, PNAS, 110(26), 10546-10551(2013).

\bibitem{Xintao15} Xintao Bi, Jeil Jung, and Zhenhua Qiao, Phys. Rev. B 92, 235421(2015).

\bibitem{Changhee16} Changhee Lee, Gunn Kim, Jeil Jung, and Hongki Min, Phys. Rev. B 94, 125438(2016).

\bibitem{YafeiRen17} Yafei Ren, Junjie Zeng, Ke Wang, Fuming Xu, and Zhenhua Qiao, Phys. Rev. B 96, 155445(2017)

\bibitem{TaoHou18} Tao Hou, Guanghui Cheng, Wang-Kong Tse, Changgan Zeng, and Zhenhua Qiao, Phys. Rev. B 98, 245417(2018).


\bibitem{Giovannetti07} Gianluca Giovannetti, Petr A. Khomyakov, Geert Brocks, Paul J. Kelly, and Jeroen van den Brink, Phys. Rev. B 76, 073103(2007)

\bibitem{CRDean10} C. R. Dean, A. F. Young, I. Meric, C. Lee, L. Wang, S. Sorgenfrei, K. Watanabe, T. Taniguchi, P. Kim, K. L. Shepard and J. Hone, Nature Nanotechnology, 5, 722每726(2010)

\bibitem{SYZhou07} S. Y. Zhou, G.-H. Gweon, A. V. Fedorov, P. N. First, W. A. de Heer, D.-H. Lee, F. Guinea, A. H. Castro Neto and A. Lanzara, Nature Materials, 6, 770每775(2007)


\bibitem{ZhenhuaQiao11} Zhenhua Qiao, Jeil Jung, Qian Niu, and Allan H. MacDonald, Nano Lett., 11 (8), 3453每3459(2011)

\bibitem{ZhenhuaQiao14} Zhenhua Qiao, Jeil Jung, Chungwei Lin, Yafei Ren, Allan H. MacDonald, and Qian Niu, Phys. Rev. Lett. 112, 206601(2014)

\bibitem{Anglin17} J. R. Anglin and A. Schulz, Phys. Rev. B 95, 045430(2017)

\bibitem{Wang17} Ke Wang, Yafei Ren, Xinzhou Deng, Shengyuan A. Yang, Jeil Jung, and Zhenhua Qiao, Phys. Rev. B 95, 245420(2017)

\bibitem{ShuguangCheng18} Shu-guang Cheng, Haiwen Liu, Hua Jiang, Qing-Feng Sun, and X. C. Xie, Phys. Rev. Lett. 121, 156801(2018)


\bibitem{Guinea10} F. Guinea, A. K. Geim, M. I. Katsnelson, and K. S. Novoselov, Phys. Rev. B 81, 035408(2010)

\bibitem{ZhigangWang13} Zhigang Wang, Zhen-Guo Fu, Fawei Zheng, and Ping Zhang, Phys. Rev. B 87, 125418(2013).

\bibitem{ZenanQi14} Zenan Qi, Alexander L. Kitt, Harold S. Park, Vitor M. Pereira, David K. Campbell, and A. H. Castro Neto, Phys. Rev. B 90, 125419(2014)

\bibitem{ShuzeZhu15} Shuze Zhu, Joseph A. Stroscio, and Teng Li, Phys. Rev. Lett. 115, 245501(2015)

\bibitem{Settnes16} Mikkel Settnes, Stephen R. Power, Mads Brandbyge, and Antti-Pekka Jauho, Phys. Rev. Lett. 117, 276801(2016)


\bibitem{LopezSancho84} M P Lopez Sancho, J M Lopez Sancho and J Rubio, Phys. F: Met. Phys., 14, 1205-1215(1984).

\bibitem{Nardelli98} Marco Buongiorno Nardelli, Phys. Rev. B, 60, 7828(1999).

\bibitem{FufangXu07} Fufang Xu, Baolei Li, Hui Pan, and Jia-Lin Zhu, Phys. Rev. B 75, 085431(2007).

\bibitem{Diniz12} G. S. Diniz, A. Latg谷, and S. E. Ulloa, Phys. Rev. Lett. 108, 126601(2012).

\bibitem{Lewenkopf13} Caio H. Lewenkopf, Eduardo R. Mucciolo, J. Comput. Electron., 12, 203每231(2013).


\bibitem{Yokoyama08} Takehito Yokoyama, Phys. Rev. B 77, 073413(2008).

\bibitem{Haugen08} Havard Haugen, Daniel Huertas-Hernando, and Arne Brataas, Phys. Rev. B 77, 115406(2008).

\bibitem{Inoue16} Takuya Inoue, Gerrit E. W. Bauer, and Kentaro Nomura, Phys. Rev. B 94, 205428(2016).




\end{thebibliography}
\end{document}